\documentclass[%
reprint,
superscriptaddress,
 amsmath,amssymb,
 aps,
 prl,
]{revtex4-2}

\usepackage{graphicx}
\usepackage{dcolumn}
\usepackage{bm}

\begin{document}


\title{The dynamical Franz-Keldysh effect in the deep ultraviolet probed by transient absorption and dispersion of diamond using a miniature beamline}

\author{Jan Reislöhner}
\affiliation{
Friedrich Schiller University Jena, Institute of Optics and Quantum Electronics, Abbe Center of Photonics, Max-Wien-Platz 1, 07743 Jena, Germany
}
\author{Xiao Chen}
\affiliation{
Friedrich Schiller University Jena, Institute of Condensed Matter Theory and Optics, Abbe Center of Photonics, Max-Wien-Platz 1, 07743 Jena, Germany
}
\author{Doyeong Kim}
\affiliation{
Friedrich Schiller University Jena, Institute of Optics and Quantum Electronics, Abbe Center of Photonics, Max-Wien-Platz 1, 07743 Jena, Germany
}
\author{Silvana Botti}
\affiliation{
Friedrich Schiller University Jena, Institute of Condensed Matter Theory and Optics, Abbe Center of Photonics, Max-Wien-Platz 1, 07743 Jena, Germany
}
\author{Adrian N. Pfeiffer}
\email{a.n.pfeiffer@uni-jena.de}
\affiliation{
Friedrich Schiller University Jena, Institute of Optics and Quantum Electronics, Abbe Center of Photonics, Max-Wien-Platz 1, 07743 Jena, Germany
}
\date{\today}

\begin{abstract} 
The deep ultraviolet, the bandgap region of dielectrics, is not readily accessible for established methods of ultrafast spectroscopy.
Here, a miniature beamline, where a tailored deep ultraviolet field is used immediately after the noncollinear generation without subsequent optical elements, is introduced for transient absorption and dispersion spectroscopy. The near-bandgap region in diamond in the presence of a few-femtosecond pump pulse is explored where the delayed dynamical Franz-Keldysh effect and the almost instantaneous optical Kerr effect coexist. 
\end{abstract}

\maketitle

Since half a century, the rapid progress of ultrafast laser technology has enabled a wealth of new technologies that are based on the manipulation of matter using light pulses. Light-induced electronic effects, which are reversible and can therefore be repeated, are suited for signal processing and communication. The extremely successful synergy of semiconductor electronics and optics is more and more plagued by bandwidth limits and heat dissipation \cite{RN306}. New technologies could be enabled by electronics in dielectrics. This idea, which seems counterintuitive at first glance, has been increasingly developed in the attosecond community since it was demonstrated that femtosecond laser pulses induce reversible charge dynamics in dielectrics at optical frequencies \cite{RN163, RN134}. This is the basis for the intriguing prospect of petahertz electronics \cite{RN132}. 

To explore the fundamental effects, numerous studies have appeared in recent years using attosecond transient absorption \cite{RN136}. Although it was recognized from the beginning that probing valence band - conduction band transitions would directly scrutinize the effects \cite{RN134}, only probe pulses with energies much higher than the bandgap have been used. The reason is that laser pulses with durations short enough to probe the electronic timescale can be generated both in the extreme ultraviolet (EUV) \cite{RN175} and in the visible-infrared (Vis-IR) spectral regions \cite{RN23}, but their generation the deep ultraviolet (DUV) remains difficult. Considerable progress in the generation of sub-5-fs DUV pulses has been demonstrated in recent years by several methods \cite{RN167, RN298, RN299}, where the shortest duration of 1.5\,fs has been achieved by concatenation of spatial harmonics that emerge from noncollinear IR-Vis pulses in a thin MgF$_2$ platelet \cite{RN194}. 

Despite advances in the generation and characterization of ultrashort DUV pulses, their application in spectroscopy is still rare. An example where DUV pulses have been used for spectroscopy, instead of only being generated and characterized, is pump-probe spectroscopy of thymine using $\sim$\,10-fs-DUV pulses \cite{RN148}. Particular challenges are the separation of the DUV field from the generating pulses and the refocusing into the sample. Optical elements, such as spectral filters and focusing mirrors, readily induce broadening and distortion before a DUV pulse reaches the sample, because optical materials in the DUV exhibit very strong absorption and dispersion.

\begin{figure*}[t]
\includegraphics[width=0.9\textwidth]{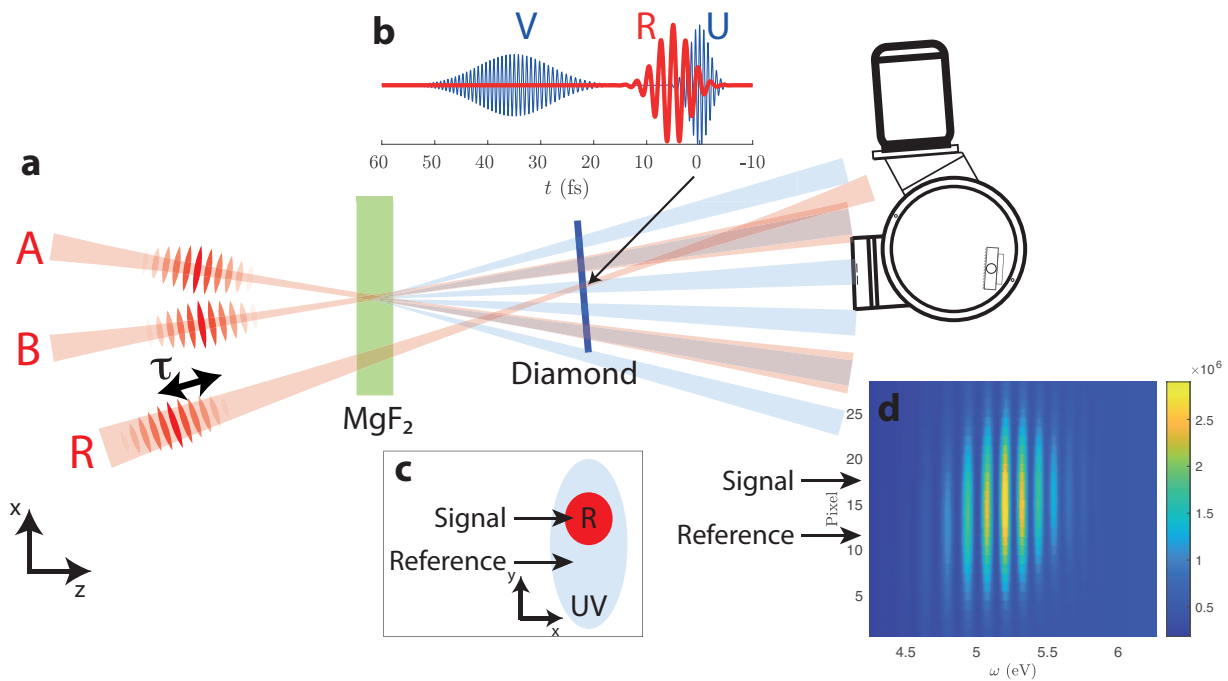}
\caption{\label{fig:setup} Miniature beamline for transient absorption and dispersion in the DUV. \textbf{a} The Vis-IR pulses A and B are used to generate the DUV-pulses U and V in the generation sample (an MgF$_2$ platelet). The TADS sample (a free-standing 250\,nm-thick diamond membrane) is placed a few centimeters behind the generation sample where it is overlapped with the pump pulse R. The polarization of all fields is in y-direction. The normalized electric fields of the pulse sequence locally at the TADS sample is shown in \textbf{b}. The transverse beam profiles of U, V and R at the location of the TADS-sample are shown in \textbf{c}. \textbf{d} The spatial resolution in the y-dimension of a DUV spectrometer is used to discriminate the signal (originating from the spatial overlap of the DUV pulses with R in \textbf{c}) and the reference (originating from a location outside the beam profile of R in \textbf{c}).}
\end{figure*}

Here, measurements of DUV transient absorption and dispersion in a free-standing sheet of diamond are presented. Unlike the traditional design consisting of separate stages for the generation, the refocusing and the probing of a sample, a novel design in a miniature beamline is introduced. The DUV field emerging from wave mixing in a non-collinear geometry is used without any subsequent optical elements. The miniature beamline is thus very compact compared to traditional designs with meter-long beamlines \cite{RN297}. The main advantage, however, is not the reduced size but that the generated fields are spatially separated from the fundamental pulses and can therefore be used without compromising the ultrashort duration. The DUV pulse sequence that interacts with a Vis-IR pump pulse in the diamond sheet, which is placed a few centimeters behind the generation sample, consists of a first DUV pulse that is very short and a second DUV pulse that is chirped. The interference of these two pulses enables the simultaneous measurement of both amplitude (transient absorption) and phase (transient dispersion). 

Using this novel type of spectroscopy, the dynamical Franz-Keldysh effect (DFKE) \cite{RN302} is scrutinized in a regime where it was deemed impossible to be observed \cite{RN300}. Previously, the DFKE has been observed only in the Vis-IR around the energy gaps of semiconductors \cite{RN309, RN304, RN308, RN303} and multiple quantum wells \cite{RN307}. The DFKE in dielectrics has recently been observed in the EUV using attosecond transient absorption but the direct probing of the band gap in the DUV was deemed impossible because of multiphoton ionization \cite{RN300}. Contrary to this expectation, it is shown here that the pump-induced changes are fully reversible and do not show a significant contribution of multiphoton absorption. By probing the regime in the proximity of the absorption edge, a previously unexplored regime is tested where both the DFKE and the optical Kerr effect (OKE) contribute to the transient absorption and dispersion spectroscopy (TADS).

In the miniature beamline, shown in Fig.\,\ref{fig:setup}, two fundamental Vis-IR pulses are focused onto a generation sample (a 100-$\mu$m-thick MgF$_2$ platelet) and give rise to a DUV field emitted from the platelet as described in Ref.\,\cite{RN194}. Counter the intuition that a single DUV pulse would be created in this process, the DUV field contains two pulses spaced 35\,fs apart, of which the first pulse U is very short and the second pulse V is chirped \cite{RN194}. The first arriving pulse U originates from the back side of the platelet, while the later arriving pulse V is generated at the front side and undergoes dispersive pulse broadening. Cross-phase modulation scans (XPMS) are used for experimental pulse retrieval \cite{RN192} and indicate $t^{FWHM}_U = 3.4$\,fs and $t^{FWHM}_V = 10.0$\,fs. The GVD is 1\,fs$^2$ for U and 12\,fs$^2$ for V. The DUV light in the miniature beamline does not consist of light pulses in the original sense but is a synthesized field with spatial-temporal couplings that exhibits ultrashort waveforms at certain emission angles. A DUV spectrometer is positioned at the emission direction where the shortest waveforms emerge. The DUV spectrum is fringed due to the interference of U and V. The TADS sample (a 250\,nm thick free-standing polycrystalline diamond sheet) is placed behind the generation sample. The experiment is carried out in vacuum. Excitation is provided by an intense Vis-IR pulse denoted R (center wavelength 750\,nm, pulse duration $t^{FWHM}_R = 5$\,fs, peak intensity $I_{R}$ = 2\,TW/cm$^2$). R is aligned such that the DUV light that is collected by the spectrometer interacts with R inside the TADS sample. The delay $\tau$ is set with a piezo-driven linear adjuster (SmarAct GmbH). 

The time resolution $\Delta t$ of the miniature beamline is limited by the pulse duration of pulse U (ideally $\Delta t = t^{FWHM}_U$) but further effects contribute. Since U and R are noncollinear (let $\beta$ denote their crossing angle), the temporal overlap depends on the transverse coordinate $x$. A simple estimation using ray optics is $\delta t_{\beta} = \frac{w_{eff}}{c}\tan(\beta)$, where $w_{eff}$ is the effective beam size of U in x-dimension and $c$ is the speed of light. The distance from the origin of the DUV radiation is 12.5\,cm to the TADS sample and 50\,cm to the spectometer slit. Thus, $w_{eff} = \frac{12.5}{50} b = 25\,\mu$m with the slit width $b = 100\,\mu$m. The angle $\beta$ should be set as small as possible, but under the condition that the pump pulse is spatially separated at the spectrometer slit. Experimentally, $\beta = 1.2\,^{\circ}$ was realized, corresponding to a time resolution of $\delta t_{\beta} = 1.7$\,fs. 
To estimate the effective time resolution, the convolution of a Gaussian pulse with $t^{FWHM}_U = 3.4\,$fs with a rectangular function of width $\Delta t_{\beta}= 1.7$\,fs is calculated. This results in $\Delta t = 3.5\,$fs, which is rather close to the limit of $t^{FWHM}_U$. 

The miniature beamline provides a reference measurement, which serves to suppress noise induced by laser power fluctuations. The beams are aligned such that a part of the DUV beam is excited by R (signal), another part is not excited (reference). To enable this, the beam geometry is adjusted such that the vertical extension of the DUV pulses (650\,$\mu$m) is significantly larger compared to the pump pulse (80\,$\mu$m) on the TADS sample (Fig.\,\ref{fig:setup}\,\textbf{c}). The spatial offset transfers from the TADS sample to the spectrometer camera, where signal and reference are separated by regions-of-interest (Fig.\,\ref{fig:setup}\,\textbf{d}). 

\begin{figure} [t]
\includegraphics[width=0.5\textwidth]{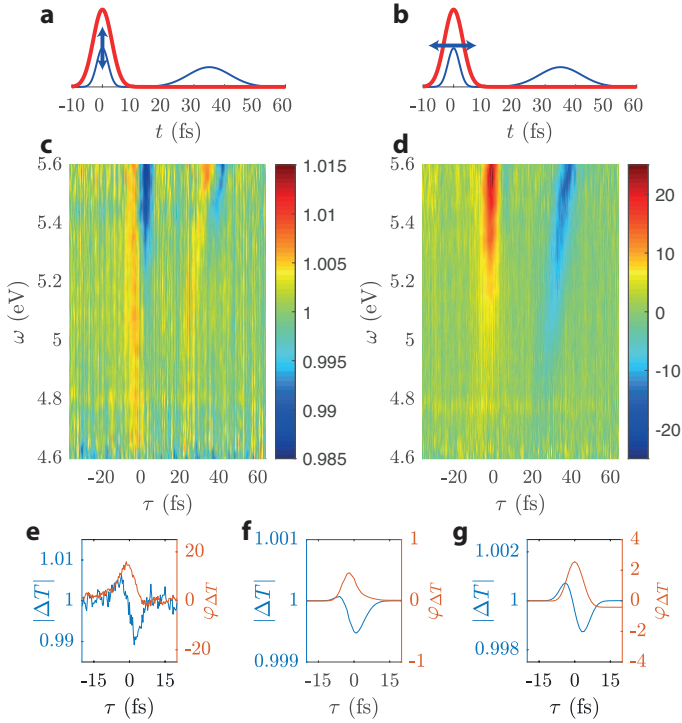}
\caption{\label{fig:data}
The sketches provide a visual interpretation of transient absorption (\textbf{a}) and transient dispersion (\textbf{b}). 
The experimental data of a 250\,nm-thick diamond membrane are shown in \textbf{c}-\textbf{e}. The magnitude $|\Delta T|$ is displayed in \textbf{c} and the phase $\varphi_{\Delta T}$ in \textbf{d}. The average in the interval $[5.3, 5,5]$\,eV is shown in \textbf{e} for comparison with the calculations using 2 bands (\textbf{f}) and 3 bands (\textbf{g}).
}
\end{figure}

To evaluate the TADS data, the complex quantity $I_+(\omega,\tau) = U^*(\omega,\tau) V(\omega,\tau)$ is used, i.e., the AC part of the double pulse spectrum. This is determined for both the signal and the reference. The transient transmittance $\Delta T$ is calculated by
\begin{align}
	\Delta T(\omega,\tau) = \frac{I_+^{sig}(\omega,\tau)}{I_+^{ref}(\omega,\tau)} \frac{I_+^{ref}(\omega,\infty)}{I_+^{sig}(\omega,\infty)}. 
	\label{eq:delta_T}
\end{align}
The data at $\tau = \infty$ correspond to the unexcited sample, where the pump pulse R reaches the TADS sample after the probe pulses U and V. The transient transmittance is a complex quantity: the magnitude $|\Delta T|$ corresponds to a conventional transient absorption measurement (sketched in Fig.\,\ref{fig:data}\,\textbf{a}). Additionally, the phase $\varphi_{\Delta T}$ is available here, which is the transient dispersion. The simplest interpretation of transient dispersion is that the pump pulse induces a time shift in the probe pulse (sketched in Fig.\,\ref{fig:data}\,\textbf{b}). Transient dispersion measurements have already been proposed for EUV experiments \cite{RN296}. 

The data displayed in Fig.\,\ref{fig:data} originate from averaging over 40 pulse delay scans, which further improves the signal-to-noise ratio in addition to the correction by the reference signal. This leads to a statistical error of $0.08\,\%$ for the magnitude $|\Delta T|$ and 0.8\,mrad for the phase $\varphi_{\Delta T}$. With an optical period time of 0.8\,fs of the DUV pulses, the phase error corresponds to 0.1\,as time error. 

\begin{figure} [t]
\includegraphics[width=0.5\textwidth]{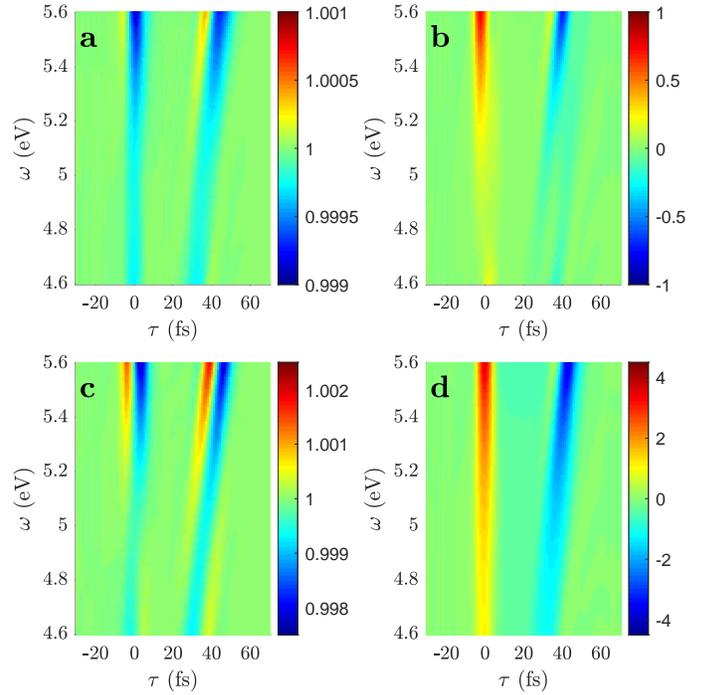}
\caption{\label{fig:sim}
The calculation of TADS in a 5\,nm-thick diamond membrane using 2 bands (\textbf{a} and \textbf{b}) and 3 bands (\textbf{c} and \textbf{d}). The magnitude $|\Delta T|$ is displayed in \textbf{a} and \textbf{c} and the phase $\varphi_{\Delta T}$ in \textbf{b} and \textbf{d}.
}
\end{figure}

\begin{figure*}[t]
\includegraphics[width=0.9\textwidth]{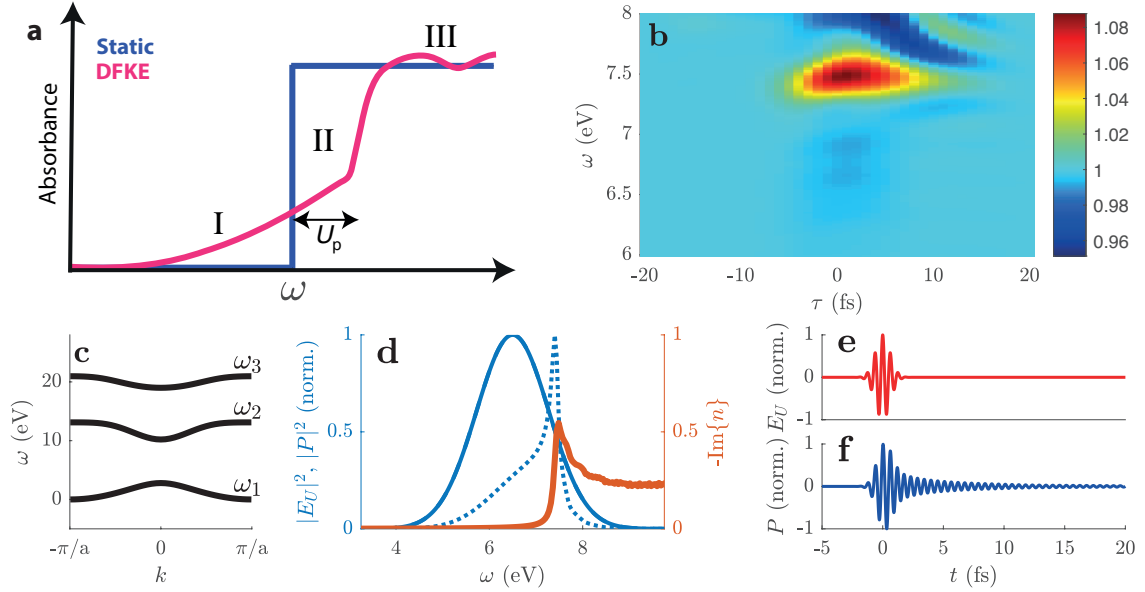}
\caption{\label{fig:FK} The DFKE. \textbf{a} The DFKE predicts absorption below the band edge (I), blueshift of the band edge by $U_p$ causing
induced transparency (II), oscillations above the band edge (III). \textbf{b} The magnitude $|\Delta T|$ calculated using 2 bands with $t^{FWHM}_U = 1$\,fs and $\omega^{center}_U = 6.5$\,eV. \textbf{c} the band energies for the calculations with 2 bands ($\omega_1$ and $\omega_2$) and 3 bands (additionally $\omega_3$); $a$ is the lattice constant. \textbf{d} The electric field of U ($E_U$, blue solid) and the linear polarization response ($P$, blue dotted) in spectral domain. The extinction coefficient (the imaginary part of the refractive index of the band model) is shown in red. The time-domain representations of $E_U$ and $P$ are shown in \textbf{e} and \textbf{f}. 
}
\end{figure*}

The overlap with the first DUV pulse U at $\tau = 0$ generates straight absorption and dispersion pattern, confirming the absence of chirp and its short pulse duration. The overlap with the second DUV pulse V around $\tau = 30$\,fs generates curved structures, indicating its chirp. The transient amplitude (Fig.\,\ref{fig:data}\,\textbf{c}) shows both enhanced absorption (blue) and enhanced transmission (red) with changes of up to 1.5\,$\%$. The transient phase (Fig.\,\ref{fig:data}\,\textbf{d}) shows changes of up to 25 mrad. No transient changes are observed in the region between the pulses U and V, which implies that the observed processes are fully reversible. The pulse delay $\tau$ is set such that $\tau=0$ is near the temporal overlap of R and U. The inspection of Fig.\,\ref{fig:data}\,\textbf{e} suggests that the transient absorption is delayed with respect to the transient dispersion by about 3\,fs.

Semiconductor Bloch equations (SBEs) are used in tandem with pulse propagation as described in the Supplemental Material. The noncollinear beam geometry must be implemented also in the calculations to avoid that third-order harmonic generation in the sample interferes with the probe pulses and obscures the TADS signal. This is a further challenge compared to the established methods in the EUV, where high-order harmonic generation in the probe sample can usually be neglected. The sample thickness is reduced to 5\,nm to decrease the computing time. The 2-band calculation, including one valence and one conduction band (Fig.\,\ref{fig:sim}\,\textbf{a} and \textbf{b}), reproduces the overall shape of the transient features, but does not reproduce the enhanced transmittance. 

The 2-band calculation is in agreement with the DFKE, which is related to the intraband motion of an electron-hole pair in the crystal potential that is bent by an laser electric field \cite{RN302}. The DFKE is an ultrafast nonresonant process in which no real carriers are created and is therefore in agreement with no persistent changes outside the temporal overlap. Transmission changes attributed to the DFKE include (Fig.\,\ref{fig:FK}\,\textbf{a}): absorption below the band edge (region I), blueshift of the band edge by approximately the ponderomotive potential $U_p$ (0.1\,eV in this study), effectively increasing the transparency (region II), and oscillatory behavior above the band edge (region III). To verify these predictions, a calculation is performed where the DUV photon energy is increased to 6.5\,eV and the pulse duration is decreased to 1\,fs (Fig.\,\ref{fig:FK}). The features of the DFKE are reproduced in all three regions. The features of the DFKE begin at pulse overlap for pump before probe ($\tau<0$), but extend to about 15\,fs when the pump arrives later ($\tau>0$). This delay on the femtosecond timescale is not to be confused with the subcycle dependence that was investigated in Refs.\,\cite{RN300, RN311, RN305}. Our interpretation of the femtosecond delay is related to the linear polarization response $P$ that is induced by the electric field $E_U$ of the probe pulse. As seen in Fig.\,\ref{fig:FK}\,\textbf{d}, the spectrum of $P$ is narrowed when $E_U$ is near the band edge, because $P(\omega) \propto \chi^{(1)}(\omega) E_U(\omega)$, where $\chi^{(1)}(\omega)$ is the susceptibility. In time domain, $P(t)$ is triggered by $E_U(t)$ but extends further for about 15\,fs (Fig.\,\ref{fig:FK}\,\textbf{e} and \textbf{f}). A later arriving pump pulse can therefore alter the polarization response although the probe field has passed. This effect has received great attention in transient absorption studies of atoms, where a later arriving pump can change the absorption line shape (e.g., \cite{RN312}). The absorption in region I, which branches into the experimentally measured region, is not as strongly delayed as the main feature in region II, but a slight delay is nevertheless present. 

Transient transmission changes in the region inside the energy gap and far below the absorption edge are usually called two-beam coupling \cite{RN313} and related to the OKE. There, the nonlinear polarization response $P^{\mathrm{NL}}$ is usually modeled as an instantaneous response $P^{(\mathrm{NL})}(t) \propto \chi^{(3)} E(t)^3$, where $\chi^{(3)}$ is the third-order susceptibility. In the present experiment, the region where the DFKE and the OKE exist on equal footing is explored. The quantitative reproduction of the OKE requires interband coupling \cite{RN249}. With the inclusion of a third band at higher energies (Fig.\,\ref{fig:FK}\,\textbf{c}), the experimental data are very well reproduced (Fig.\,\ref{fig:sim}\,\textbf{c} and \textbf{d}). The inclusion of the third band affects $\varphi_{\Delta T}$ much more than $|\Delta T|$, which is clearly seen in Fig.\,\ref{fig:data}\,\textbf{f} and \textbf{g}. This finally explains the experimental observation that transient absorption (caused by DFKE and OKE) is delayed with respect to the transient dispersion (mainly caused by OKE). 

Diamond was proven to be a good candidate for optical information-processing on a chip \cite{RN310}. Experiments using the miniature beamline give insight in bandgap dynamics and their implications for optical transmission changes, which will hopefully stimulate integrated nonlinear optics covering a wide wavelength range and finally approaching petahertz frequencies.

\begin{acknowledgments}
This project was supported primarily by the Deutsche Forschungsgemeinschaft (DFG, German Research Foundation) - project ID 398816777 -
via projects A2 and B1 in the Collaborative Research Centre 1375 "Nonlinear optics down to atomic scales" (NOA). DK is funded via DFG Priority Programme 1840 - project ID 281272215 - "Quantum Dynamics in Tailored Intense Fields" (QUTIF). 

JR and ANP conducted the experiment. ANP provided the calculations using SBEs. XC and SB supported band structure calculations. All authors contributed to the discussion and preparation of the manuscript.
\end{acknowledgments}


\begin{thebibliography}{28}%
\makeatletter
\providecommand \@ifxundefined [1]{%
 \@ifx{#1\undefined}
}%
\providecommand \@ifnum [1]{%
 \ifnum #1\expandafter \@firstoftwo
 \else \expandafter \@secondoftwo
 \fi
}%
\providecommand \@ifx [1]{%
 \ifx #1\expandafter \@firstoftwo
 \else \expandafter \@secondoftwo
 \fi
}%
\providecommand \natexlab [1]{#1}%
\providecommand \enquote  [1]{``#1''}%
\providecommand \bibnamefont  [1]{#1}%
\providecommand \bibfnamefont [1]{#1}%
\providecommand \citenamefont [1]{#1}%
\providecommand \href@noop [0]{\@secondoftwo}%
\providecommand \href [0]{\begingroup \@sanitize@url \@href}%
\providecommand \@href[1]{\@@startlink{#1}\@@href}%
\providecommand \@@href[1]{\endgroup#1\@@endlink}%
\providecommand \@sanitize@url [0]{\catcode `\\12\catcode `\$12\catcode
  `\&12\catcode `\#12\catcode `\^12\catcode `\_12\catcode `\%12\relax}%
\providecommand \@@startlink[1]{}%
\providecommand \@@endlink[0]{}%
\providecommand \url  [0]{\begingroup\@sanitize@url \@url }%
\providecommand \@url [1]{\endgroup\@href {#1}{\urlprefix }}%
\providecommand \urlprefix  [0]{URL }%
\providecommand \Eprint [0]{\href }%
\providecommand \doibase [0]{https://doi.org/}%
\providecommand \selectlanguage [0]{\@gobble}%
\providecommand \bibinfo  [0]{\@secondoftwo}%
\providecommand \bibfield  [0]{\@secondoftwo}%
\providecommand \translation [1]{[#1]}%
\providecommand \BibitemOpen [0]{}%
\providecommand \bibitemStop [0]{}%
\providecommand \bibitemNoStop [0]{.\EOS\space}%
\providecommand \EOS [0]{\spacefactor3000\relax}%
\providecommand \BibitemShut  [1]{\csname bibitem#1\endcsname}%
\let\auto@bib@innerbib\@empty
\bibitem [{\citenamefont {Caulfield}\ and\ \citenamefont
  {Dolev}(2010)}]{RN306}%
  \BibitemOpen
  \bibfield  {author} {\bibinfo {author} {\bibfnamefont {H.~J.}\ \bibnamefont
  {Caulfield}}\ and\ \bibinfo {author} {\bibfnamefont {S.}~\bibnamefont
  {Dolev}},\ }\bibfield  {title} {\bibinfo {title} {Why future supercomputing
  requires optics},\ }\href@noop {} {\bibfield  {journal} {\bibinfo  {journal}
  {Nature Photonics}\ }\textbf {\bibinfo {volume} {4}},\ \bibinfo {pages} {261}
  (\bibinfo {year} {2010})}\BibitemShut {NoStop}%
\bibitem [{\citenamefont {Schiffrin}\ \emph {et~al.}(2013)\citenamefont
  {Schiffrin}, \citenamefont {Paasch-Colberg}, \citenamefont {Karpowicz},
  \citenamefont {Apalkov}, \citenamefont {Gerster}, \citenamefont {Muhlbrandt},
  \citenamefont {Korbman}, \citenamefont {Reichert}, \citenamefont {Schultze},
  \citenamefont {Holzner}, \citenamefont {Barth}, \citenamefont {Kienberger},
  \citenamefont {Ernstorfer}, \citenamefont {Yakovlev}, \citenamefont
  {Stockman},\ and\ \citenamefont {Krausz}}]{RN163}%
  \BibitemOpen
  \bibfield  {author} {\bibinfo {author} {\bibfnamefont {A.}~\bibnamefont
  {Schiffrin}}, \bibinfo {author} {\bibfnamefont {T.}~\bibnamefont
  {Paasch-Colberg}}, \bibinfo {author} {\bibfnamefont {N.}~\bibnamefont
  {Karpowicz}}, \bibinfo {author} {\bibfnamefont {V.}~\bibnamefont {Apalkov}},
  \bibinfo {author} {\bibfnamefont {D.}~\bibnamefont {Gerster}}, \bibinfo
  {author} {\bibfnamefont {S.}~\bibnamefont {Muhlbrandt}}, \bibinfo {author}
  {\bibfnamefont {M.}~\bibnamefont {Korbman}}, \bibinfo {author} {\bibfnamefont
  {J.}~\bibnamefont {Reichert}}, \bibinfo {author} {\bibfnamefont
  {M.}~\bibnamefont {Schultze}}, \bibinfo {author} {\bibfnamefont
  {S.}~\bibnamefont {Holzner}}, \bibinfo {author} {\bibfnamefont {J.~V.}\
  \bibnamefont {Barth}}, \bibinfo {author} {\bibfnamefont {R.}~\bibnamefont
  {Kienberger}}, \bibinfo {author} {\bibfnamefont {R.}~\bibnamefont
  {Ernstorfer}}, \bibinfo {author} {\bibfnamefont {V.~S.}\ \bibnamefont
  {Yakovlev}}, \bibinfo {author} {\bibfnamefont {M.~I.}\ \bibnamefont
  {Stockman}},\ and\ \bibinfo {author} {\bibfnamefont {F.}~\bibnamefont
  {Krausz}},\ }\bibfield  {title} {\bibinfo {title} {Optical-field-induced
  current in dielectrics},\ }\href@noop {} {\bibfield  {journal} {\bibinfo
  {journal} {Nature}\ }\textbf {\bibinfo {volume} {493}},\ \bibinfo {pages}
  {70} (\bibinfo {year} {2013})}\BibitemShut {NoStop}%
\bibitem [{\citenamefont {Schultze}\ \emph {et~al.}(2013)\citenamefont
  {Schultze}, \citenamefont {Bothschafter}, \citenamefont {Sommer},
  \citenamefont {Holzner}, \citenamefont {Schweinberger}, \citenamefont
  {Fiess}, \citenamefont {Hofstetter}, \citenamefont {Kienberger},
  \citenamefont {Apalkov}, \citenamefont {Yakovlev}, \citenamefont {Stockman},\
  and\ \citenamefont {Krausz}}]{RN134}%
  \BibitemOpen
  \bibfield  {author} {\bibinfo {author} {\bibfnamefont {M.}~\bibnamefont
  {Schultze}}, \bibinfo {author} {\bibfnamefont {E.~M.}\ \bibnamefont
  {Bothschafter}}, \bibinfo {author} {\bibfnamefont {A.}~\bibnamefont
  {Sommer}}, \bibinfo {author} {\bibfnamefont {S.}~\bibnamefont {Holzner}},
  \bibinfo {author} {\bibfnamefont {W.}~\bibnamefont {Schweinberger}}, \bibinfo
  {author} {\bibfnamefont {M.}~\bibnamefont {Fiess}}, \bibinfo {author}
  {\bibfnamefont {M.}~\bibnamefont {Hofstetter}}, \bibinfo {author}
  {\bibfnamefont {R.}~\bibnamefont {Kienberger}}, \bibinfo {author}
  {\bibfnamefont {V.}~\bibnamefont {Apalkov}}, \bibinfo {author} {\bibfnamefont
  {V.~S.}\ \bibnamefont {Yakovlev}}, \bibinfo {author} {\bibfnamefont {M.~I.}\
  \bibnamefont {Stockman}},\ and\ \bibinfo {author} {\bibfnamefont
  {F.}~\bibnamefont {Krausz}},\ }\bibfield  {title} {\bibinfo {title}
  {Controlling dielectrics with the electric field of light},\ }\href@noop {}
  {\bibfield  {journal} {\bibinfo  {journal} {Nature}\ }\textbf {\bibinfo
  {volume} {493}},\ \bibinfo {pages} {75} (\bibinfo {year} {2013})}\BibitemShut
  {NoStop}%
\bibitem [{\citenamefont {Krausz}\ and\ \citenamefont
  {Stockman}(2014)}]{RN132}%
  \BibitemOpen
  \bibfield  {author} {\bibinfo {author} {\bibfnamefont {F.}~\bibnamefont
  {Krausz}}\ and\ \bibinfo {author} {\bibfnamefont {M.~I.}\ \bibnamefont
  {Stockman}},\ }\bibfield  {title} {\bibinfo {title} {Attosecond metrology:
  from electron capture to future signal processing},\ }\href@noop {}
  {\bibfield  {journal} {\bibinfo  {journal} {Nature Photonics}\ }\textbf
  {\bibinfo {volume} {8}},\ \bibinfo {pages} {205} (\bibinfo {year}
  {2014})}\BibitemShut {NoStop}%
\bibitem [{\citenamefont {Borja}\ \emph {et~al.}(2016)\citenamefont {Borja},
  \citenamefont {Zurch}, \citenamefont {Pemmaraju}, \citenamefont {Schultze},
  \citenamefont {Ramasesha}, \citenamefont {Gandman}, \citenamefont {Prell},
  \citenamefont {Prendergast}, \citenamefont {Neumark},\ and\ \citenamefont
  {Leone}}]{RN136}%
  \BibitemOpen
  \bibfield  {author} {\bibinfo {author} {\bibfnamefont {L.~J.}\ \bibnamefont
  {Borja}}, \bibinfo {author} {\bibfnamefont {M.}~\bibnamefont {Zurch}},
  \bibinfo {author} {\bibfnamefont {C.~D.}\ \bibnamefont {Pemmaraju}}, \bibinfo
  {author} {\bibfnamefont {M.}~\bibnamefont {Schultze}}, \bibinfo {author}
  {\bibfnamefont {K.}~\bibnamefont {Ramasesha}}, \bibinfo {author}
  {\bibfnamefont {A.}~\bibnamefont {Gandman}}, \bibinfo {author} {\bibfnamefont
  {J.~S.}\ \bibnamefont {Prell}}, \bibinfo {author} {\bibfnamefont
  {D.}~\bibnamefont {Prendergast}}, \bibinfo {author} {\bibfnamefont {D.~M.}\
  \bibnamefont {Neumark}},\ and\ \bibinfo {author} {\bibfnamefont {S.~R.}\
  \bibnamefont {Leone}},\ }\bibfield  {title} {\bibinfo {title} {Extreme
  ultraviolet transient absorption of solids from femtosecond to attosecond
  timescales [invited]},\ }\href@noop {} {\bibfield  {journal} {\bibinfo
  {journal} {Journal of the Optical Society of America B-Optical Physics}\
  }\textbf {\bibinfo {volume} {33}},\ \bibinfo {pages} {C57} (\bibinfo {year}
  {2016})}\BibitemShut {NoStop}%
\bibitem [{\citenamefont {Krausz}\ and\ \citenamefont {Ivanov}(2009)}]{RN175}%
  \BibitemOpen
  \bibfield  {author} {\bibinfo {author} {\bibfnamefont {F.}~\bibnamefont
  {Krausz}}\ and\ \bibinfo {author} {\bibfnamefont {M.}~\bibnamefont
  {Ivanov}},\ }\bibfield  {title} {\bibinfo {title} {Attosecond physics},\
  }\href@noop {} {\bibfield  {journal} {\bibinfo  {journal} {Reviews of Modern
  Physics}\ }\textbf {\bibinfo {volume} {81}},\ \bibinfo {pages} {163}
  (\bibinfo {year} {2009})}\BibitemShut {NoStop}%
\bibitem [{\citenamefont {Manzoni}\ \emph {et~al.}(2015)\citenamefont
  {Manzoni}, \citenamefont {Mucke}, \citenamefont {Cirmi}, \citenamefont
  {Fang}, \citenamefont {Moses}, \citenamefont {Huang}, \citenamefont {Hong},
  \citenamefont {Cerullo},\ and\ \citenamefont {Kartner}}]{RN23}%
  \BibitemOpen
  \bibfield  {author} {\bibinfo {author} {\bibfnamefont {C.}~\bibnamefont
  {Manzoni}}, \bibinfo {author} {\bibfnamefont {O.~D.}\ \bibnamefont {Mucke}},
  \bibinfo {author} {\bibfnamefont {G.}~\bibnamefont {Cirmi}}, \bibinfo
  {author} {\bibfnamefont {S.~B.}\ \bibnamefont {Fang}}, \bibinfo {author}
  {\bibfnamefont {J.}~\bibnamefont {Moses}}, \bibinfo {author} {\bibfnamefont
  {S.~W.}\ \bibnamefont {Huang}}, \bibinfo {author} {\bibfnamefont {K.~H.}\
  \bibnamefont {Hong}}, \bibinfo {author} {\bibfnamefont {G.}~\bibnamefont
  {Cerullo}},\ and\ \bibinfo {author} {\bibfnamefont {F.~X.}\ \bibnamefont
  {Kartner}},\ }\bibfield  {title} {\bibinfo {title} {Coherent pulse synthesis:
  towards sub-cycle optical waveforms},\ }\href@noop {} {\bibfield  {journal}
  {\bibinfo  {journal} {Laser and Photonics Reviews}\ }\textbf {\bibinfo
  {volume} {9}},\ \bibinfo {pages} {129} (\bibinfo {year} {2015})}\BibitemShut
  {NoStop}%
\bibitem [{\citenamefont {Reiter}\ \emph {et~al.}(2010)\citenamefont {Reiter},
  \citenamefont {Graf}, \citenamefont {Serebryannikov}, \citenamefont
  {Schweinberger}, \citenamefont {Fiess}, \citenamefont {Schultze},
  \citenamefont {Azzeer}, \citenamefont {Kienberger}, \citenamefont {Krausz},
  \citenamefont {Zheltikov},\ and\ \citenamefont {Goulielmakis}}]{RN167}%
  \BibitemOpen
  \bibfield  {author} {\bibinfo {author} {\bibfnamefont {F.}~\bibnamefont
  {Reiter}}, \bibinfo {author} {\bibfnamefont {U.}~\bibnamefont {Graf}},
  \bibinfo {author} {\bibfnamefont {E.~E.}\ \bibnamefont {Serebryannikov}},
  \bibinfo {author} {\bibfnamefont {W.}~\bibnamefont {Schweinberger}}, \bibinfo
  {author} {\bibfnamefont {M.}~\bibnamefont {Fiess}}, \bibinfo {author}
  {\bibfnamefont {M.}~\bibnamefont {Schultze}}, \bibinfo {author}
  {\bibfnamefont {A.~M.}\ \bibnamefont {Azzeer}}, \bibinfo {author}
  {\bibfnamefont {R.}~\bibnamefont {Kienberger}}, \bibinfo {author}
  {\bibfnamefont {F.}~\bibnamefont {Krausz}}, \bibinfo {author} {\bibfnamefont
  {A.~M.}\ \bibnamefont {Zheltikov}},\ and\ \bibinfo {author} {\bibfnamefont
  {E.}~\bibnamefont {Goulielmakis}},\ }\bibfield  {title} {\bibinfo {title}
  {Route to attosecond nonlinear spectroscopy},\ }\href@noop {} {\bibfield
  {journal} {\bibinfo  {journal} {Phys Rev Lett}\ }\textbf {\bibinfo {volume}
  {105}},\ \bibinfo {pages} {243902} (\bibinfo {year} {2010})}\BibitemShut
  {NoStop}%
\bibitem [{\citenamefont {Galli}\ \emph {et~al.}(2019)\citenamefont {Galli},
  \citenamefont {Wanie}, \citenamefont {Lopes}, \citenamefont {Mansson},
  \citenamefont {Trabattoni}, \citenamefont {Colaizzi}, \citenamefont
  {Saraswathula}, \citenamefont {Cartella}, \citenamefont {Frassetto},
  \citenamefont {Poletto}, \citenamefont {Legare}, \citenamefont {Stagira},
  \citenamefont {Nisoli}, \citenamefont {Vazquez}, \citenamefont {Osellame},\
  and\ \citenamefont {Calegari}}]{RN298}%
  \BibitemOpen
  \bibfield  {author} {\bibinfo {author} {\bibfnamefont {M.}~\bibnamefont
  {Galli}}, \bibinfo {author} {\bibfnamefont {V.}~\bibnamefont {Wanie}},
  \bibinfo {author} {\bibfnamefont {D.~P.}\ \bibnamefont {Lopes}}, \bibinfo
  {author} {\bibfnamefont {E.~P.}\ \bibnamefont {Mansson}}, \bibinfo {author}
  {\bibfnamefont {A.}~\bibnamefont {Trabattoni}}, \bibinfo {author}
  {\bibfnamefont {L.}~\bibnamefont {Colaizzi}}, \bibinfo {author}
  {\bibfnamefont {K.}~\bibnamefont {Saraswathula}}, \bibinfo {author}
  {\bibfnamefont {A.}~\bibnamefont {Cartella}}, \bibinfo {author}
  {\bibfnamefont {F.}~\bibnamefont {Frassetto}}, \bibinfo {author}
  {\bibfnamefont {L.}~\bibnamefont {Poletto}}, \bibinfo {author} {\bibfnamefont
  {F.}~\bibnamefont {Legare}}, \bibinfo {author} {\bibfnamefont
  {S.}~\bibnamefont {Stagira}}, \bibinfo {author} {\bibfnamefont
  {M.}~\bibnamefont {Nisoli}}, \bibinfo {author} {\bibfnamefont {R.~M.}\
  \bibnamefont {Vazquez}}, \bibinfo {author} {\bibfnamefont {R.}~\bibnamefont
  {Osellame}},\ and\ \bibinfo {author} {\bibfnamefont {F.}~\bibnamefont
  {Calegari}},\ }\bibfield  {title} {\bibinfo {title} {Generation of deep
  ultraviolet sub-2-fs pulses},\ }\href@noop {} {\bibfield  {journal} {\bibinfo
   {journal} {Optics Letters}\ }\textbf {\bibinfo {volume} {44}},\ \bibinfo
  {pages} {1308} (\bibinfo {year} {2019})}\BibitemShut {NoStop}%
\bibitem [{\citenamefont {Travers}\ \emph {et~al.}(2019)\citenamefont
  {Travers}, \citenamefont {Grigorova}, \citenamefont {Brahms},\ and\
  \citenamefont {Belli}}]{RN299}%
  \BibitemOpen
  \bibfield  {author} {\bibinfo {author} {\bibfnamefont {J.~C.}\ \bibnamefont
  {Travers}}, \bibinfo {author} {\bibfnamefont {T.~F.}\ \bibnamefont
  {Grigorova}}, \bibinfo {author} {\bibfnamefont {C.}~\bibnamefont {Brahms}},\
  and\ \bibinfo {author} {\bibfnamefont {F.}~\bibnamefont {Belli}},\ }\bibfield
   {title} {\bibinfo {title} {High-energy pulse self-compression and
  ultraviolet generation through soliton dynamics in hollow capillary fibres},\
  }\href@noop {} {\bibfield  {journal} {\bibinfo  {journal} {Nature Photonics}\
  }\textbf {\bibinfo {volume} {13}},\ \bibinfo {pages} {547} (\bibinfo {year}
  {2019})}\BibitemShut {NoStop}%
\bibitem [{\citenamefont {Reislöhner}\ \emph
  {et~al.}(2019{\natexlab{a}})\citenamefont {Reislöhner}, \citenamefont
  {Leithold},\ and\ \citenamefont {Pfeiffer}}]{RN194}%
  \BibitemOpen
  \bibfield  {author} {\bibinfo {author} {\bibfnamefont {J.}~\bibnamefont
  {Reislöhner}}, \bibinfo {author} {\bibfnamefont {C.}~\bibnamefont
  {Leithold}},\ and\ \bibinfo {author} {\bibfnamefont {A.~N.}\ \bibnamefont
  {Pfeiffer}},\ }\bibfield  {title} {\bibinfo {title} {Harmonic concatenation
  of 1.5 fs pulses in the deep ultraviolet},\ }\href@noop {} {\bibfield
  {journal} {\bibinfo  {journal} {ACS Photonics}\ }\textbf {\bibinfo {volume}
  {6}},\ \bibinfo {pages} {1351} (\bibinfo {year}
  {2019}{\natexlab{a}})}\BibitemShut {NoStop}%
\bibitem [{\citenamefont {Kobayashi}\ and\ \citenamefont {Kida}(2012)}]{RN148}%
  \BibitemOpen
  \bibfield  {author} {\bibinfo {author} {\bibfnamefont {T.}~\bibnamefont
  {Kobayashi}}\ and\ \bibinfo {author} {\bibfnamefont {Y.}~\bibnamefont
  {Kida}},\ }\bibfield  {title} {\bibinfo {title} {Ultrafast spectroscopy with
  sub-10 fs deep-ultraviolet pulses},\ }\href@noop {} {\bibfield  {journal}
  {\bibinfo  {journal} {Phys Chem Chem Phys}\ }\textbf {\bibinfo {volume}
  {14}},\ \bibinfo {pages} {6200} (\bibinfo {year} {2012})}\BibitemShut
  {NoStop}%
\bibitem [{\citenamefont {Beck}\ \emph {et~al.}(2015)\citenamefont {Beck},
  \citenamefont {Neumark},\ and\ \citenamefont {Leone}}]{RN297}%
  \BibitemOpen
  \bibfield  {author} {\bibinfo {author} {\bibfnamefont {A.~R.}\ \bibnamefont
  {Beck}}, \bibinfo {author} {\bibfnamefont {D.~M.}\ \bibnamefont {Neumark}},\
  and\ \bibinfo {author} {\bibfnamefont {S.~R.}\ \bibnamefont {Leone}},\
  }\bibfield  {title} {\bibinfo {title} {Probing ultrafast dynamics with
  attosecond transient absorption},\ }\href@noop {} {\bibfield  {journal}
  {\bibinfo  {journal} {Chemical Physics Letters}\ }\textbf {\bibinfo {volume}
  {624}},\ \bibinfo {pages} {119} (\bibinfo {year} {2015})}\BibitemShut
  {NoStop}%
\bibitem [{\citenamefont {Jauho}\ and\ \citenamefont {Johnsen}(1996)}]{RN302}%
  \BibitemOpen
  \bibfield  {author} {\bibinfo {author} {\bibfnamefont {A.~P.}\ \bibnamefont
  {Jauho}}\ and\ \bibinfo {author} {\bibfnamefont {K.}~\bibnamefont
  {Johnsen}},\ }\bibfield  {title} {\bibinfo {title} {Dynamical franz-keldysh
  effect},\ }\href@noop {} {\bibfield  {journal} {\bibinfo  {journal} {Physical
  Review Letters}\ }\textbf {\bibinfo {volume} {76}},\ \bibinfo {pages} {4576}
  (\bibinfo {year} {1996})}\BibitemShut {NoStop}%
\bibitem [{\citenamefont {Lucchini}\ \emph {et~al.}(2016)\citenamefont
  {Lucchini}, \citenamefont {Sato}, \citenamefont {Ludwig}, \citenamefont
  {Herrmann}, \citenamefont {Volkov}, \citenamefont {Kasmi}, \citenamefont
  {Shinohara}, \citenamefont {Yabana}, \citenamefont {Gallmann},\ and\
  \citenamefont {Keller}}]{RN300}%
  \BibitemOpen
  \bibfield  {author} {\bibinfo {author} {\bibfnamefont {M.}~\bibnamefont
  {Lucchini}}, \bibinfo {author} {\bibfnamefont {S.~A.}\ \bibnamefont {Sato}},
  \bibinfo {author} {\bibfnamefont {A.}~\bibnamefont {Ludwig}}, \bibinfo
  {author} {\bibfnamefont {J.}~\bibnamefont {Herrmann}}, \bibinfo {author}
  {\bibfnamefont {M.}~\bibnamefont {Volkov}}, \bibinfo {author} {\bibfnamefont
  {L.}~\bibnamefont {Kasmi}}, \bibinfo {author} {\bibfnamefont
  {Y.}~\bibnamefont {Shinohara}}, \bibinfo {author} {\bibfnamefont
  {K.}~\bibnamefont {Yabana}}, \bibinfo {author} {\bibfnamefont
  {L.}~\bibnamefont {Gallmann}},\ and\ \bibinfo {author} {\bibfnamefont
  {U.}~\bibnamefont {Keller}},\ }\bibfield  {title} {\bibinfo {title}
  {Attosecond dynamical franz-keldysh effect in polycrystalline diamond},\
  }\href@noop {} {\bibfield  {journal} {\bibinfo  {journal} {Science}\ }\textbf
  {\bibinfo {volume} {353}},\ \bibinfo {pages} {916} (\bibinfo {year}
  {2016})}\BibitemShut {NoStop}%
\bibitem [{\citenamefont {Chin}\ \emph {et~al.}(2000)\citenamefont {Chin},
  \citenamefont {Bakker},\ and\ \citenamefont {Kono}}]{RN309}%
  \BibitemOpen
  \bibfield  {author} {\bibinfo {author} {\bibfnamefont {A.~H.}\ \bibnamefont
  {Chin}}, \bibinfo {author} {\bibfnamefont {J.~M.}\ \bibnamefont {Bakker}},\
  and\ \bibinfo {author} {\bibfnamefont {J.}~\bibnamefont {Kono}},\ }\bibfield
  {title} {\bibinfo {title} {Ultrafast electroabsorption at the transition
  between classical and quantum response},\ }\href@noop {} {\bibfield
  {journal} {\bibinfo  {journal} {Physical Review Letters}\ }\textbf {\bibinfo
  {volume} {85}},\ \bibinfo {pages} {3293} (\bibinfo {year}
  {2000})}\BibitemShut {NoStop}%
\bibitem [{\citenamefont {Srivastava}\ \emph {et~al.}(2004)\citenamefont
  {Srivastava}, \citenamefont {Srivastava}, \citenamefont {Wang},\ and\
  \citenamefont {Kono}}]{RN304}%
  \BibitemOpen
  \bibfield  {author} {\bibinfo {author} {\bibfnamefont {A.}~\bibnamefont
  {Srivastava}}, \bibinfo {author} {\bibfnamefont {R.}~\bibnamefont
  {Srivastava}}, \bibinfo {author} {\bibfnamefont {J.~G.}\ \bibnamefont
  {Wang}},\ and\ \bibinfo {author} {\bibfnamefont {J.}~\bibnamefont {Kono}},\
  }\bibfield  {title} {\bibinfo {title} {Laser-induced above-band-gap
  transparency in gaas},\ }\href@noop {} {\bibfield  {journal} {\bibinfo
  {journal} {Physical Review Letters}\ }\textbf {\bibinfo {volume} {93}}
  (\bibinfo {year} {2004})}\BibitemShut {NoStop}%
\bibitem [{\citenamefont {Ghimire}\ \emph {et~al.}(2011)\citenamefont
  {Ghimire}, \citenamefont {DiChiara}, \citenamefont {Sistrunk}, \citenamefont
  {Szafruga}, \citenamefont {Agostini}, \citenamefont {DiMauro},\ and\
  \citenamefont {Reis}}]{RN308}%
  \BibitemOpen
  \bibfield  {author} {\bibinfo {author} {\bibfnamefont {S.}~\bibnamefont
  {Ghimire}}, \bibinfo {author} {\bibfnamefont {A.~D.}\ \bibnamefont
  {DiChiara}}, \bibinfo {author} {\bibfnamefont {E.}~\bibnamefont {Sistrunk}},
  \bibinfo {author} {\bibfnamefont {U.~B.}\ \bibnamefont {Szafruga}}, \bibinfo
  {author} {\bibfnamefont {P.}~\bibnamefont {Agostini}}, \bibinfo {author}
  {\bibfnamefont {L.~F.}\ \bibnamefont {DiMauro}},\ and\ \bibinfo {author}
  {\bibfnamefont {D.~A.}\ \bibnamefont {Reis}},\ }\bibfield  {title} {\bibinfo
  {title} {Redshift in the optical absorption of zno single crystals in the
  presence of an intense midinfrared laser field},\ }\href@noop {} {\bibfield
  {journal} {\bibinfo  {journal} {Physical Review Letters}\ }\textbf {\bibinfo
  {volume} {107}} (\bibinfo {year} {2011})}\BibitemShut {NoStop}%
\bibitem [{\citenamefont {Novelli}\ \emph {et~al.}(2013)\citenamefont
  {Novelli}, \citenamefont {Fausti}, \citenamefont {Giusti}, \citenamefont
  {Parmigiani},\ and\ \citenamefont {Hoffmann}}]{RN303}%
  \BibitemOpen
  \bibfield  {author} {\bibinfo {author} {\bibfnamefont {F.}~\bibnamefont
  {Novelli}}, \bibinfo {author} {\bibfnamefont {D.}~\bibnamefont {Fausti}},
  \bibinfo {author} {\bibfnamefont {F.}~\bibnamefont {Giusti}}, \bibinfo
  {author} {\bibfnamefont {F.}~\bibnamefont {Parmigiani}},\ and\ \bibinfo
  {author} {\bibfnamefont {M.}~\bibnamefont {Hoffmann}},\ }\bibfield  {title}
  {\bibinfo {title} {Mixed regime of light-matter interaction revealed by phase
  sensitive measurements of the dynamical franz-keldysh effect},\ }\href@noop
  {} {\bibfield  {journal} {\bibinfo  {journal} {Scientific Reports}\ }\textbf
  {\bibinfo {volume} {3}} (\bibinfo {year} {2013})}\BibitemShut {NoStop}%
\bibitem [{\citenamefont {Nordstrom}\ \emph {et~al.}(1998)\citenamefont
  {Nordstrom}, \citenamefont {Johnsen}, \citenamefont {Allen}, \citenamefont
  {Jauho}, \citenamefont {Birnir}, \citenamefont {Kono}, \citenamefont {Noda},
  \citenamefont {Akiyama},\ and\ \citenamefont {Sakaki}}]{RN307}%
  \BibitemOpen
  \bibfield  {author} {\bibinfo {author} {\bibfnamefont {K.~B.}\ \bibnamefont
  {Nordstrom}}, \bibinfo {author} {\bibfnamefont {K.}~\bibnamefont {Johnsen}},
  \bibinfo {author} {\bibfnamefont {S.~J.}\ \bibnamefont {Allen}}, \bibinfo
  {author} {\bibfnamefont {A.~P.}\ \bibnamefont {Jauho}}, \bibinfo {author}
  {\bibfnamefont {B.}~\bibnamefont {Birnir}}, \bibinfo {author} {\bibfnamefont
  {J.}~\bibnamefont {Kono}}, \bibinfo {author} {\bibfnamefont {T.}~\bibnamefont
  {Noda}}, \bibinfo {author} {\bibfnamefont {H.}~\bibnamefont {Akiyama}},\ and\
  \bibinfo {author} {\bibfnamefont {H.}~\bibnamefont {Sakaki}},\ }\bibfield
  {title} {\bibinfo {title} {Excitonic dynamical franz-keldysh effect},\
  }\href@noop {} {\bibfield  {journal} {\bibinfo  {journal} {Physical Review
  Letters}\ }\textbf {\bibinfo {volume} {81}},\ \bibinfo {pages} {457}
  (\bibinfo {year} {1998})}\BibitemShut {NoStop}%
\bibitem [{\citenamefont {Reislöhner}\ \emph
  {et~al.}(2019{\natexlab{b}})\citenamefont {Reislöhner}, \citenamefont
  {Leithold},\ and\ \citenamefont {Pfeiffer}}]{RN192}%
  \BibitemOpen
  \bibfield  {author} {\bibinfo {author} {\bibfnamefont {J.}~\bibnamefont
  {Reislöhner}}, \bibinfo {author} {\bibfnamefont {C.}~\bibnamefont
  {Leithold}},\ and\ \bibinfo {author} {\bibfnamefont {A.~N.}\ \bibnamefont
  {Pfeiffer}},\ }\bibfield  {title} {\bibinfo {title} {Characterization of weak
  deep uv pulses using cross-phase modulation scans},\ }\href@noop {}
  {\bibfield  {journal} {\bibinfo  {journal} {Opt. Lett.}\ }\textbf {\bibinfo
  {volume} {44}},\ \bibinfo {pages} {1809} (\bibinfo {year}
  {2019}{\natexlab{b}})}\BibitemShut {NoStop}%
\bibitem [{\citenamefont {Pfeifer}\ \emph {et~al.}(2008)\citenamefont
  {Pfeifer}, \citenamefont {Abel}, \citenamefont {Nagel}, \citenamefont
  {Jullien}, \citenamefont {Loh}, \citenamefont {Bell}, \citenamefont
  {Neumark},\ and\ \citenamefont {Leone}}]{RN296}%
  \BibitemOpen
  \bibfield  {author} {\bibinfo {author} {\bibfnamefont {T.}~\bibnamefont
  {Pfeifer}}, \bibinfo {author} {\bibfnamefont {M.~J.}\ \bibnamefont {Abel}},
  \bibinfo {author} {\bibfnamefont {P.~M.}\ \bibnamefont {Nagel}}, \bibinfo
  {author} {\bibfnamefont {A.}~\bibnamefont {Jullien}}, \bibinfo {author}
  {\bibfnamefont {Z.~H.}\ \bibnamefont {Loh}}, \bibinfo {author} {\bibfnamefont
  {M.~J.}\ \bibnamefont {Bell}}, \bibinfo {author} {\bibfnamefont {D.~M.}\
  \bibnamefont {Neumark}},\ and\ \bibinfo {author} {\bibfnamefont {S.~R.}\
  \bibnamefont {Leone}},\ }\bibfield  {title} {\bibinfo {title} {Time-resolved
  spectroscopy of attosecond quantum dynamics},\ }\href@noop {} {\bibfield
  {journal} {\bibinfo  {journal} {Chemical Physics Letters}\ }\textbf {\bibinfo
  {volume} {463}},\ \bibinfo {pages} {11} (\bibinfo {year} {2008})}\BibitemShut
  {NoStop}%
\bibitem [{\citenamefont {Otobe}\ \emph {et~al.}(2016)\citenamefont {Otobe},
  \citenamefont {Shinohara}, \citenamefont {Sato},\ and\ \citenamefont
  {Yabana}}]{RN311}%
  \BibitemOpen
  \bibfield  {author} {\bibinfo {author} {\bibfnamefont {T.}~\bibnamefont
  {Otobe}}, \bibinfo {author} {\bibfnamefont {Y.}~\bibnamefont {Shinohara}},
  \bibinfo {author} {\bibfnamefont {S.~A.}\ \bibnamefont {Sato}},\ and\
  \bibinfo {author} {\bibfnamefont {K.}~\bibnamefont {Yabana}},\ }\bibfield
  {title} {\bibinfo {title} {Femtosecond time-resolved dynamical franz-keldysh
  effect},\ }\href@noop {} {\bibfield  {journal} {\bibinfo  {journal} {Physical
  Review B}\ }\textbf {\bibinfo {volume} {93}} (\bibinfo {year}
  {2016})}\BibitemShut {NoStop}%
\bibitem [{\citenamefont {Lucchini}\ \emph {et~al.}(2020)\citenamefont
  {Lucchini}, \citenamefont {Sato}, \citenamefont {Schlaepfer}, \citenamefont
  {Yabana}, \citenamefont {Gallmann}, \citenamefont {Rubio},\ and\
  \citenamefont {Keller}}]{RN305}%
  \BibitemOpen
  \bibfield  {author} {\bibinfo {author} {\bibfnamefont {M.}~\bibnamefont
  {Lucchini}}, \bibinfo {author} {\bibfnamefont {S.~A.}\ \bibnamefont {Sato}},
  \bibinfo {author} {\bibfnamefont {F.}~\bibnamefont {Schlaepfer}}, \bibinfo
  {author} {\bibfnamefont {K.}~\bibnamefont {Yabana}}, \bibinfo {author}
  {\bibfnamefont {L.}~\bibnamefont {Gallmann}}, \bibinfo {author}
  {\bibfnamefont {A.}~\bibnamefont {Rubio}},\ and\ \bibinfo {author}
  {\bibfnamefont {U.}~\bibnamefont {Keller}},\ }\bibfield  {title} {\bibinfo
  {title} {Attosecond timing of the dynamical franz-keldysh effect},\
  }\href@noop {} {\bibfield  {journal} {\bibinfo  {journal} {Journal of
  Physics-Photonics}\ }\textbf {\bibinfo {volume} {2}} (\bibinfo {year}
  {2020})}\BibitemShut {NoStop}%
\bibitem [{\citenamefont {Ott}\ \emph {et~al.}(2013)\citenamefont {Ott},
  \citenamefont {Kaldun}, \citenamefont {Raith}, \citenamefont {Meyer},
  \citenamefont {Laux}, \citenamefont {Evers}, \citenamefont {Keitel},
  \citenamefont {Greene},\ and\ \citenamefont {Pfeifer}}]{RN312}%
  \BibitemOpen
  \bibfield  {author} {\bibinfo {author} {\bibfnamefont {C.}~\bibnamefont
  {Ott}}, \bibinfo {author} {\bibfnamefont {A.}~\bibnamefont {Kaldun}},
  \bibinfo {author} {\bibfnamefont {P.}~\bibnamefont {Raith}}, \bibinfo
  {author} {\bibfnamefont {K.}~\bibnamefont {Meyer}}, \bibinfo {author}
  {\bibfnamefont {M.}~\bibnamefont {Laux}}, \bibinfo {author} {\bibfnamefont
  {J.}~\bibnamefont {Evers}}, \bibinfo {author} {\bibfnamefont {C.~H.}\
  \bibnamefont {Keitel}}, \bibinfo {author} {\bibfnamefont {C.~H.}\
  \bibnamefont {Greene}},\ and\ \bibinfo {author} {\bibfnamefont
  {T.}~\bibnamefont {Pfeifer}},\ }\bibfield  {title} {\bibinfo {title} {Lorentz
  meets fano in spectral line shapes: A universal phase and its laser
  control},\ }\href@noop {} {\bibfield  {journal} {\bibinfo  {journal}
  {Science}\ }\textbf {\bibinfo {volume} {340}},\ \bibinfo {pages} {716}
  (\bibinfo {year} {2013})}\BibitemShut {NoStop}%
\bibitem [{\citenamefont {Smolorz}\ and\ \citenamefont {Wise}(2000)}]{RN313}%
  \BibitemOpen
  \bibfield  {author} {\bibinfo {author} {\bibfnamefont {S.}~\bibnamefont
  {Smolorz}}\ and\ \bibinfo {author} {\bibfnamefont {F.}~\bibnamefont {Wise}},\
  }\bibfield  {title} {\bibinfo {title} {Femtosecond two-beam coupling energy
  transfer from raman and electronic nonlinearities},\ }\href@noop {}
  {\bibfield  {journal} {\bibinfo  {journal} {Journal of the Optical Society of
  America B-Optical Physics}\ }\textbf {\bibinfo {volume} {17}},\ \bibinfo
  {pages} {1636} (\bibinfo {year} {2000})}\BibitemShut {NoStop}%
\bibitem [{\citenamefont {Pfeiffer}(2020)}]{RN249}%
  \BibitemOpen
  \bibfield  {author} {\bibinfo {author} {\bibfnamefont {A.~N.}\ \bibnamefont
  {Pfeiffer}},\ }\bibfield  {title} {\bibinfo {title} {Iteration of
  semiconductor bloch equations for ultrashort laser pulse propagation},\
  }\href@noop {} {\bibfield  {journal} {\bibinfo  {journal} {Journal of Physics
  B-Atomic Molecular and Optical Physics}\ }\textbf {\bibinfo {volume} {53}}
  (\bibinfo {year} {2020})}\BibitemShut {NoStop}%
\bibitem [{\citenamefont {Hausmann}\ \emph {et~al.}(2014)\citenamefont
  {Hausmann}, \citenamefont {Bulu}, \citenamefont {Venkataraman}, \citenamefont
  {Deotare},\ and\ \citenamefont {Loncar}}]{RN310}%
  \BibitemOpen
  \bibfield  {author} {\bibinfo {author} {\bibfnamefont {B.~J.~M.}\
  \bibnamefont {Hausmann}}, \bibinfo {author} {\bibfnamefont {I.}~\bibnamefont
  {Bulu}}, \bibinfo {author} {\bibfnamefont {V.}~\bibnamefont {Venkataraman}},
  \bibinfo {author} {\bibfnamefont {P.}~\bibnamefont {Deotare}},\ and\ \bibinfo
  {author} {\bibfnamefont {M.}~\bibnamefont {Loncar}},\ }\bibfield  {title}
  {\bibinfo {title} {Diamond nonlinear photonics},\ }\href@noop {} {\bibfield
  {journal} {\bibinfo  {journal} {Nature Photonics}\ }\textbf {\bibinfo
  {volume} {8}},\ \bibinfo {pages} {369} (\bibinfo {year} {2014})}\BibitemShut
  {NoStop}%
\end{thebibliography}

%

\end{document}